\documentclass[11pt]{article}
\title{Quantum phenomenology and the Continuum Problem }
\author{O. Yaremchuk\footnote{email: yarem@sci.lebedev.ru}}
\date{}
\begin{document}
\maketitle
\begin{abstract}
A new approach to quantum mechanics based on independence of the
Continuum Hypothesis is proposed. In one-dimensional case, it is
shown that the properties of the set of intermediate cardinality
coincide with quantum phenomenology.

\smallskip
\noindent PACS nubers: 03.65.Bz, 02.10.Cz

\end{abstract}

The concept of discrete space is always regarded as the unique alternative of
the continuous space. Nevertheless, it is important to stress that there is the
intermediate possibility connected with the Continuum Problem (discrete space
is a countable set). According to the independence of the Continuum Hypothesis (CH) 
we can neither prove nor disprove existence of a set of cardinality between
cardinalities of countably infinite set and continuum. If Zermelo-Fraenkel
set theory is consistent and complete, then the uncertain status of the
intermediate set is unavoidable: no one can state that the set does not exist
but at the same time  we in principle can not get it anyhow. In other words,
uncertainty seems to be a property of the intermediate set. On the other hand,
this set standing midway between countable set and continuum may combine
properties of continuity (wave) and countability (particle). Hence,
Wave-Particle duality may be considered as the direct pointing to the
Continuum Problem and the intermediate set.

Suppose there exists a set $S$ such that $$card(N)<card(S)<card(R),$$ where $N$ 
is the set of natural numbers, $R$ is the set of all real numbers (let $S$ be 
called an interset). Then $R$ contains a subset $M$ equivalent to $S$, i.e.,
there exists a bijection $f:S\to M=f(S)\subset R$ separating $M$ from
the set of real numbers. But it follows from independence of CH that the interset
can not be separated from continuum: any procedure of separation is a proof of
the negation of the Continuum Hypothesis. Hence, for any real number $r\in R$ we do 
not know, in principle, if the sentence $r\in M$ is true or false. Therefore, any
bijection can not take a point $s\in S$ to a definite real number (the uncertain
status of the intermediate set causes uncertainty of its members with
respect to real numbers).

Then the point $s$  corresponds to entire continuum $R$ and we can
assign to the point a function $\psi(r)$ defined on the same domain $R$.
If a map takes $s$  to a random real number, then the function
$\psi(r)$ has to be connected with probability $P(r)$ of finding the point at
$r$: $P(r)={\cal P}(\psi (r))\Delta r$.
For  of two real numbers $a$ and $b$ probability $P_{a\cup b}$
of finding the point in the union of their neighbourhoods
$P_{a\cup b}\ne P(a)+P(b)$ because the point $s$ corresponds to both
intervals at the same time (the events are not mutually exclusive).
Then it is convenient to choose the function $\psi (r)$ such that 
$$\psi_{a\cup b}=\psi(a)+\psi(b).$$ 
But 
$$P_{a\cup b}={\cal P}(\psi_{a\cup b})\Delta r={\cal P}(\psi(a)+\psi(b))\Delta r\ne
{\cal P}(\psi(a))\Delta r+{\cal}P(\psi(b))\Delta r.$$
Therefore,
$${\cal P}(\psi(a)+\psi(b))\ne{\cal P}(\psi(a))+{\cal}P(\psi(b)),$$
i.e., the dependence ${\cal P}(\psi (r))$ is nonlinear. The simplest nonlinear dependence
is a square dependence: $${\cal P}(r)=|\psi (r)|^2.$$

\medskip
 CH is under discussion for more than one hundred years but no one can 
find in the set theory literature any description of probable properties of the 
interset. It is due to the following reason. According to the separation axiom
schema for any set $X$ and for any property expressed by formula $\varphi$ there
exists a subset of the set $X$, which contains only members of $X$ having
$\varphi:$
$$\forall X\exists Y\forall u(u\in Y\leftrightarrow u\in X\wedge\varphi(u)).$$
If the interset can not be separated from $R$, then its members
have not any distinction with respect to the members of continuum
(uncertainty is not a set theory property). Hence, the properties of the 
interset are

(1) uncertainty with respect to continuum,

(2) separate properties of countable sets and continuum,

(3) undecidable equivalents of the negation of CH.

Only these properties do not contradict the independence of CH because
they do not allow separation of the interset from continuum in 
accordance with the separation axiom schema. Item (2) may be called
N-R duality. Moreover, we claim that any description of
the intermediate set must be based on combining separate
properties of the countable set and continuum.

This is the reason why the set theorists do not like to discuss probable 
properties of the interset: in some sense, it has no its own properties. It 
may be stated that this is the cause of independence of CH. If the interset 
had any unique property, then it would be possible to prove its existence.

The sets of natural and real numbers have certain structures, which may be called
natural. In consequence of dual nature of the intermediate set, its structure
must be a combination of the natural structures of the sets of natural and
real numbers. Let us form such a combination.

Any interval of continuum is equivalent to any other interval and to 
entire continuum. In other words, any arbitrarily small interval contains
the same number of points as the set of all real numbers. This property is
too strong for any real set. We will consider this property as a distinctive
feature of continuum. And for the  interset we shall substitute the
following combination of the properties of the continuous and countable
sets for the property of continuum: there exists a unit (minimal)
set (unlike continuum) but this set is infinite (unlike the set of natural
numbers). Consequently, different intervals of the interset are not equivalent
(different intervals of real numbers, generally speaking, contain subsets of
different intermediate cardinalities). Cardinality of the unit set is
an infinite fundamental constant.

Coordinate of a point in the interset in units of
the minimal set is a natural number $n$. The function
$\psi$, necessarily, depends on $n$: $$\psi(r)\to\psi(n,r).$$
Since $n$ is accurate up to a constant (shift) and the function
$\psi$ is defined up to the factor $e^{i\mbox{const}}$, we have
$$\psi (n+\mbox{const})=e^{i\mbox{const}}\psi (n).$$
Hence, the function $\psi$ is of the following form:
$$\psi (r,n)=A(r)e^{2\pi in}=A(r)e^{2\pi i\nu t},$$
where natural number $\nu$  is the (constant) time rate of
change of cardinality (number of the unit sets per second).

Thus De Broglie wave represents the simplest map from the 
interset to continuum. This means that we can replace
Wave-Particle duality by the mathematical concept.

In accordance with the dual description of the interset, 
cardinality of an interval of the interset depends on its size.
Then a sufficiently large interval approach continuum. Thus the
function $\psi$ describes the real relationship between the levels
(scales) of the intermediate set, i.e., the properties of the set
of intermediate cardinality coincide with quantum phenomenology.

The intermediate set is absolutely new set of stepwise changing 
infinite cardinality (and consequently, changing structure). In fact, 
this is a spectrum of sets. The countable set and continuum are the 
ends of the spectrum. Existence of infinite quantum of cardinality is
a unique property of the set. 

\end{document}